\documentclass[journal]{IEEEtran}
\ifCLASSINFOpdf
\usepackage[pdftex]{graphicx}
\else
\fi
\usepackage{amsmath}
\usepackage{color}
\usepackage{multirow}
\usepackage{array}
\usepackage{algpseudocode}
\usepackage{algorithm}
\usepackage{algorithmicx}
\usepackage{algpseudocode}
\usepackage{amsmath}
\usepackage{listings}

\usepackage{amsmath}
\usepackage{amssymb}
\usepackage{amsthm}
\usepackage{graphicx}
\usepackage{cite}
\usepackage{algorithm}
\usepackage{algpseudocode}

\begin{document}

%
\title{ A Canonical-based NPN Boolean Matching Algorithm Utilizing Boolean Difference and Cofactor Signature}
%
%
%

\author{Juling Zhang,
        Guowu Yang,
        William N. N. Hung,
        and Jinzhao Wu         
\thanks{J.L. Zhang, G.W. Yang are with Big Data Research Center, School of Computer Science and Engineering, University of Electronic Science and Technology of China, Chengdu, 611731, China. E-mail: (zjlgj@163.com, guowu@uestc.edu.cn).}
\thanks{W.N.N. Hung is with Synopsys Inc., Mountain View, CA, USA. E-mail: William.Hung@synopsys.com.}
\thanks{J.Z. Wu (corresponding author) is with Guangxi Key Laboratory of Hybrid Computation and IC Design Analysis, Guangxi University for Nationalities, Nanning, 530006, China. E-mail: gxmdwjzh@aliyun.com}}

%
%

\markboth{Journal of \LaTeX\ Class Files,~Vol.~14, No.~8,  November~2016}%
{Shell \MakeLowercase{\textit{et al.}}: Bare Demo of IEEEtran.cls for IEEE Journals}
%



\maketitle

\begin{abstract}
This paper presents a new compact canonical-based algorithm to solve the problem of single-output completely specified NPN Boolean matching. We propose a new signature vector Boolean difference and cofactor (DC) signature vector. Our algorithm utilizes the Boolean difference, cofactor signature and symmetry properties to search for canonical transformations. The use of symmetry and Boolean difference notably reduces the search space and speeds up the Boolean matching process compared to the algorithm proposed in \cite{Adbollahi2008}. We tested our algorithm on a large number of circuits. The experimental results showed that the average runtime of our algorithm 37\% higher and its average search space 67\% smaller compared to \cite{Adbollahi2008} when tested on general circuits.

\end{abstract}

\begin{IEEEkeywords}
Boolean Matching, Boolean difference, NPN equivalent, Symmetry.
\end{IEEEkeywords}

%
\IEEEpeerreviewmaketitle

\section{Introduction}
%
%
%
%

Judging whether two Boolean functions are equivalent under input negation and/or input permutation and/or output negation (NPN) is an important problem applied in integrated circuit design, logic synthesis , logic verification, and so on. In cryptography, affine equivalence is used to resolve S-box problem \cite{YanZhang2016}. Technology mapping is the process of selecting logic gates from a library to implement a Boolean circuit \cite{B.Kapoor11995,Mailhot1990,M.Damiani2003}. However, every Boolean function has many NPN-equivalent Boolean functions. Technology mapping searches for an optimal combination of logic gates in terms of area, performance and power dissipation. In the cell-library binding process, some of the cells of a library are found to realize some part of a multiple-level representation of a Boolean function \cite{Debnath2004}. Two Boolean functions $f$ and $g$ are NPN equivalent if a transformation $T$ exists that can transform $f$ to $g$. In technology mapping and cell-library binding, Boolean matching is a key step.

In recent years, numerous methods have emerged to solve the NPN Boolean matching problem. The main approaches focus on the canonical-based, pairwise and SAT-based algorithms. In the canonical-based algorithms, two Boolean functions are NPN-equivalent when they have the same canonical form. An NPN-equivalent class has a canonical representative, and all the Boolean functions in an NPN-equivalent class can be transformed to this canonical representative. The canonical representative may have a maximal truth table or a minimal truth table or a maximal signature vector \cite{Adbollahi2008,Agosta2009,Chai2006,Debnath2004,Ciric2003,Agosta2007,Huang2013,Petkovska2016}.

A pairwise Boolean matching algorithm searches the correspondence relation of the variables of two Boolean functions by their signatures. In the pairwise Boolean matching process, the search operation terminated when it finds a transformation $T$ that can transform one Boolean function into the other. The authors of \cite{K.C.Chen1993,Abdollahi2008,Y.T.Lai1992,Juling} presented a pairwise Boolean matching algorithm.

The NPN Boolean matching problem is converted to one that involves solving Boolean satisfiability (SAT) problem. To check whether a Boolean function $f$ is equivalent to a Boolean function $g$ , the SAT-based algorithm first constructs a circuit with the functionality $f \oplus g$ and then generates a SAT instance (circuit CNF). Finally, it judges the Boolean satisfiability of this circuit CNF \cite{WangKuoHua2009}. When the circuit CNF is satisfied, $f$ is not NPN-equivalent to $g$. In contrast, $f$ is NPN-equivalent to $g$ when the circuit CNF is not satisfied.

It is well known that the search space complexity of NPN Boolean matching is $n!2^{n+1}$ using the exhaustive method. Therefore, the exhaustive method is computationally infeasible when the number of inputs is large, and thus, it is important to reduce the search space regardless of what approach is used. Many properties of Boolean functions are used to reduce the search space, including positive unate, negative unate, cofactor and symmetry, and so on.

This paper studies canonical-based NPN Boolean matching for single-output specified Boolean functions. We exploit a new compact canonical form using a DC signature vector. The canonical representative of an NPN equivalent class has the maximal DC signature vector. The use of the symmetry class, independent class and Boolean difference signature dramatically reduce the search space of the canonical transformation. The experimental results reflect that using the Boolean difference signature is highly conductive for distinguishing variables. The proposed algorithm enhances the NPN Boolean matching speed and reduces the search space significantly.

The remainder of this paper is organized as follows. In Section \uppercase\expandafter{\romannumeral2}, we survey the state of the Boolean matching problem. Section \uppercase\expandafter{\romannumeral3} introduces related preliminaries and definitions used in our algorithm. Section \uppercase\expandafter{\romannumeral4} presents our canonical-based algorithm in detail. We demonstrate the effectiveness of our algorithm by presenting experimental results in Section \uppercase\expandafter{\romannumeral5}, and we summarize our work and outline future work in Section \uppercase\expandafter{\romannumeral6}.

\section{Related Works}
Numerous scholars have contributed to this problem in prior Boolean matching studies. The authors of \cite{Adbollahi2008,Agosta2009,Chai2006,Debnath2004,Ciric2003,Agosta2007,Huang2013,Petkovska2016} studied the canonical-based Boolean matching algorithm and proposed many practical methods for computing the canonical representative. The authors of \cite{Agosta2009} studied the P-equivalence problem and proposed a formal framework that unified the spectral and canonical-based approaches. They found a linear transformation to improve the speed of P-equivalence matching. Chai and Kuehlmann of \cite{Chai2006} designed a fast Boolean matching that used satisfied counts to assign the phases of the Boolean function and its variables, the sums of rows or columns to search permutation, and symmetry to refine transformations. In \cite{Debnath2004}, a canonical-based algorithm was presented that used table look-ups and a tree-based breadth-first search to find the canonical representative with the smallest value in the binary representation. Ciric and Sechen \cite{Ciric2003} proposed the minimum-cost canonical form for P-equivalence matching. In \cite{Agosta2007}, the authors unified multiple canonical-based approaches and proposed a new P-equivalence matching algorithm. Generally, in the canonical-based Boolean matching algorithms, the canonical representative has the maximal or minimal truth table. Abdollahi and Pedram \cite{Adbollahi2008} exploited a new canonical form using signature vector to resolve the NPN Boolean matching problem. The algorithms proposed in \cite{Huang2013} and \cite{Petkovska2016} used general symmetry and higher-order symmetry to realize fast NPN Boolean classification. They classified a number of Boolean functions with 6-16 inputs.

Many Boolean matching approaches have emerged based on SAT. In a SAT-based Boolean matching algorithm, Boolean matching is converted to a SAT or UNSAT problem. The authors of \cite{Katebi2010,Matsunaga2016,WangXiuQin2013,WangKuoHua2009,Safarpour2006,KuoHua2007} all studied Boolean matching using the SAT technique. The authors of \cite{Katebi2010} proposed PP-equivalence matching based on graphs, simulation and SAT; their approach can be applied to large-scale circuits. Matsunaga \cite{Matsunaga2016} presented a Boolean matching method for LUT-based circuits by using one-hot encoding and the CEGAR technique to speed up Boolean matching \cite{Matsunaga2016}. The authors of \cite{WangXiuQin2013} exploited architectural symmetry in PLB and proposed a SAT-based Boolean matching algorithm to resolve FPGA technology mapping. Wang et al. \cite{WangKuoHua2009} integrated the simulation and SAT technique, and proposed a P-equivalent Boolean matching approach.

 A pairwise Boolean matching algorithm was presented in \cite{K.C.Chen1993} for multiple-output Boolean functions with don't care sets and applied Boolean matching to technology mapping to optimize circuits. Abdollahi \cite{Abdollahi2008} proposed a signature-based Boolean matching to resolve NPN-equivalent problem for single-output incompletely specified Boolean functions. They tested their Boolean matching approach on 4-10 inputs Boolean functions. The authors of \cite{Y.T.Lai1992} utilized a level-first strategy and a set of filters to reduce the search space in the pairwise Boolean matching process, while \cite{Juling} unified the structural signature of Boolean functions and Shannon expansion to exploit a pairwise Boolean matching algorithm.

A few other methods exist for solving the Boolean matching problem. In \cite{Moore2008}, the authors studied NPN Boolean matching using a Walsh Spectral Decision Diagram, and achieved an efficiency higher than that of Luks' hypergraph method. Lai et al. \cite{Chih-Fan2010} utilized a conflict-driven learning method to solve the multiple-output Boolean function matching problem.

\section{Preliminaries and Problem Statement}

Let $f(x_1,x_2,\cdots,x_{n})$ be a single-output completely specified Boolean function and $X=(x_1,\cdots,x_n)$ be a vector of $f$. $|f|$ denotes the number of minterms in $f$, which is also called the 0th-order signature.

 $X_{\pi}(x_1,\cdots,x_n)=(x_{\pi(1)},\cdots,x_{\pi(n)})$ is an input permutation and $X_{\varphi}(x_1,\cdots,x_n)=(x_{1}^{\varphi(1)},\cdots,x_{n}^{\varphi(n)})$ is an input negation. $x_{i}^{\varphi(i)}=x_i$ when ${\varphi(i)}=1$, and $x_{i}^{\varphi(i)}={\overline{x}}_i$ when ${\varphi(i)}=0$. NPN equivalence is introduced by $f \equiv g \Leftrightarrow f = {(g \circ {X_\pi } \circ {X_\varphi })^\phi }$, where $\varphi \in {\cal B}^n(\phi \in {\cal B})$ \cite{Hinsberger1998}.

\textbf{Definition 1: (NP transformation)} An NP transformation, $T$, is an onto mapping where $T =\{ {x_1}, \cdots ,{x_n}\}  \mapsto \{ x_{_{\pi (1)}}^{\phi (1)}, \cdots ,x_{_{\pi (n)}}^{\phi (n)}\}  $.

For each Boolean function $f$ in an NPN-equivalent class with a canonical representative $\mathbb F$, there must be an NP transformation $\cal T$ that can transform $f$ to $\mathbb F$, (i.e., $f({\cal T}X)=\mathbb F$ or $\overline{f({\cal T}X)}=\mathbb F$). The $\cal T$ is called a canonical transformation. Boolean function $f_1$ is NPN-equivalent to $f_2$ if and only if they have the same canonical representative (i.e., ${\mathbb F}_1={\mathbb F}_2$).

A cofactor is a generalized signature of a Boolean function that has been used in many Boolean matching algorithms.

\textbf{Definition 2: (Cofactor signature):} The cofactor signature of $f$ with respect to $x_i({\overline{x}}_i)$, $|f_{x_i}|(|f_{{\overline{x}}_i}|)$, is the onset size of $f_{x_i}(f_{{\overline{x}}_i})$, where $f_{x_i}=f[x_i \leftarrow 1]$ and $f_{{\overline{x}}_i}=f[x_i \leftarrow 0]$.

The cofactor signature of $f$ with respect to a cube $b$, $|f_b|$, is the onset size of $f_b$, where $f_b=f[b \leftarrow 1]$. If the cube $b$ has $k$ variables, $|f_b|$ is denoted as the kth-order signature \cite{Adbollahi2008}.

A Boolean function may have some independent variables; however, the value of a Boolean function is irrelevant to these independent variables. The negative cofactor and positive cofactor of a variable are used to check its independence.

\textbf{Definition 3: (Boolean difference signature)} The Boolean difference signature of $f$ with respect to $x_i$,$|f_{x_i}^{'}|$, is the onset size of $f_{x_i}^{'}$, where $f_{x_i}^{'}=f_{x_i} \oplus f_{{\overline{x}}_i}$ \cite{Jin2006}.

A variable $x_i$ is said to be an independent variable when it satisfies $|{f'_{x_i}}|=0$. The independence property of an independent variable does not changed under NP transformation.

\textbf{Definition 4: (Variable symmetry):} The variables $x_i$ and $x_j({\overline{x}}_j)$ of a Boolean function $f$ are symmetric if $f$ is invariant after swapping $x_i$ and $x_j({\overline{x}}_j)$ \cite{Jin2006}.

Two NP transformations are equal when two or more symmetric variables are permuted. Therefore, symmetry is always used to reduce the search space for Boolean matching.

\section{Canonical-based Boolean Matching}
The goal of our algorithm is to find the canonical transformation $\cal T$ of a Boolean function $f$, which can transform $f$ to $\mathbb F$ that has the maximal DC signature vector. During the search process, we search many candidate canonical transformations by sorting and grouping variables. Then, we compare the DC signature vectors of these candidate canonical transformations to find the target canonical transformation $\cal T$.

\subsection{The Proposed Canonical Form}
 The canonical form presented in \cite{Adbollahi2008} has the maximal signature vector. For some Boolean functions, many variables have the same cofactor signature and are not symmetric variables. In this case, the algorithm proposed in \cite{Adbollahi2008} will generate multiple splitting ways, which increases the search space of canonical transformation. Therefore, to reduce the search space, further methods to distinguish these variables are required.

Although the variables of many Boolean functions have the same cofactor signature, their Boolean difference signatures are the different. Therefore, we proposes DC signature vector that can resolve the shortcoming in \cite{Adbollahi2008} described above.

 \textbf{Definition 5: ($\prec $)} Let "$ \prec $" denote the lexicographic comparison of two vectors.

\textbf{Definition 6: ($1^{st}$ DC signature value)} The $1^{st}$ Boolean Difference and Cofactor (DC) signature value of a Boolean function $f$ with respect to $x_i$ is a two-tuple $(|f_{x_i}|,|f_{x_i}^{'}|)$.

The $1^{st}$ DC signature values of the two variables $x_i$ and $x_j$ of a Boolean function $f$ are $(|f_{x_i}|,|f_{x_i}^{'}|)$ and $(|f_{x_j}|,|f_{x_j}^{'}|)$, respectively. When they satisfy one of the two following cases, then $(|f_{x_i}|,|f_{x_i}^{'}|) \prec (|f_{x_j}|,|f_{x_j}^{'}|)$ and the relation of $x_i$ and $x_j$ is $x_i < x_j$.

1)  $|f_{x_i}|<|f_{x_j}|$

2) $|f_{x_i}|=|f_{x_j}|$ and $|f_{x_i}^{'}|<|f_{x_j}^{'}|$

\textbf{Definition 7: (\emph{k}th-order DC signature value)} The \emph{k}th-order DC signature value of a Boolean function $f$ with respect to a cube b, $b=x_{i1}\cdots x_{ik}$, is $(|f_{x_{i1}\cdots x_{ik}}|,|f_{x_{i1}\cdots x_{ik}}^{'}|)$.

\textbf{Definition 8: ( DC signature vector)} The DC signature vector of an \emph{n}-input Boolean function $f$ is

${D^f} = \{ |f|,(|f_{x_1}|,|f_{x_1}^{'}|),\cdots,(|f_{x_n}|,|f_{x_n}^{'}|),(|f_{x_1x_2}|,|f_{x_1x_2}^{'}|),\newline \cdots,(|f_{x_{n-1}x_n}|,|f_{x_{n-1}x_n}^{'}|),\cdots,
(|f_{x_1 \cdots x_{n-1}}|,|f_{x_1 \cdots x_{n-1}}^{'}|),\newline (|f_{x_2 \cdots x_{n}}|,|f_{x_2 \cdots x_{n}}^{'}|),|f_{x_1 \cdots x_n}|\} $, which is composed of its \emph{0}th-order signature, $1^{st}$ DC signature values and higher order DC signature values up to the \emph{n}th-order signature.

The authors of \cite{Adbollahi2008} proved that each Boolean function has a unique signature vector. Therefore, each Boolean function also has a unique DC signature vector. We need not prove that here.

\textbf{Definition 9: (The canonical representative)} The canonical representative of an NPN-equivalent class $EC=\{f_1,\cdots,f_m\}$ is the Boolean function that has the maximal DC signature vector.

Assume that two Boolean functions $f$ and $g$ have the DC signature vectors $D^f$ and $D^g$. When comparing $D^f$ and $D^g$, we compare their 0th-order signature first, then their $1^{st}$ DC signature values, 2nd-order DC signature values and higher DC signature values until an inequality is encountered. $D^f \prec D^g$ when the corresponding DC signature value of $D^f$ is less than that of $D^g$. If and only if $D^f \prec D^g$, the order relation of $f$ and $g$ is $f<g$.

\textbf{Definition 10: (The maximal canonical transformation)} The transformation $C_f=(t_1,t_2,\cdots,t_n)$ that has the maximal DC signature vector is the maximal canonical transformation.

In this paper, we use $T=(t_1,t_2,\cdots,t_n)$ to denote a candidate canonical transformation searched in a canonized process where $t_i=x_{{\pi}(i)}^{{\varphi}(i)}$. For a candidate canonical transformation $T$, the DC signature vector is as follows:

${D^T} = \{ |f|,(|f_{t_1}|,|f_{t_1}^{'}|),\cdots,(|f_{t_n}|,|f_{t_n}^{'}|),(|f_{t_1t_2}|,|f_{t_1t_2}^{'}|),\newline \cdots,(|f_{t_{n-1}t_n}|,|f_{t_{n-1}t_n}^{'}|),\cdots,
(|f_{t_1 \cdots t_{n-1}}|,|f_{t_1 \cdots t_{n-1}}^{'}|),\newline (|f_{t_2 \cdots t_{n}}|,|f_{t_2 \cdots t_{n}}^{'}|),|f_{t_1 \cdots t_n}|\} $

 In the canonized process, multiple candidate canonical transformations similar to $T$ exist. The maximal canonical transformation $C_f$ is the candidate canonical transformation that has the maximal DC signature vector. Consider two candidate canonical transformations $T1$ and $T2$: $T1 < T2$ when $D^{T1} \prec D^{T2}$.

 For an \emph{n}-input Boolean function $f$, $n!2^n$ NP transformations exist. The canonical transformation $\cal T$ of a Boolean function $f$ is the transformation that can transform $f$ to $\mathbb F$ when $\mathbb F$ has the maximal DC signature vector in its NPN-equivalent class. We use $C_f$ to express the maximal canonical transformation of Boolean function $f$ found in the canonical process. The input phase assignment and reindexing transformation is $X=(x_1,\cdots,x_n)$ to $C_f=(t_1,\cdots,t_n)$, where ${\mathbb F}(C_f)=f(X)$. The relation between $f$ and $\mathbb F$ can be expressed by an NP transformation $\cal T$, where $X={\cal T}C_f$ \cite{Adbollahi2008}. Therefore, the canonical form of a Boolean function $f$ is $f({\cal T}X)$, and $f({\cal T}X)=f({C_f}^{-1}X)$.

 The key step of our algorithm is sorting and grouping the variables according to their DC signature values. After each grouping, a group result is obtained. We denote $G$ as the group result of $f$, and $G=\{G_1,\cdots,G_m\}$, where $m \in \{1,\cdots,n\}$. Every group $G_i$ has one or multiple classes, and each class has one or multiple variables. We use $G_i=\{C_{i1},\cdots,C_{ik}\}$ denote a group with $k$ classes.

 In the process of grouping variables, each group may have many classes. We categorize these as asymmetric, symmetric and independent classes.

 \textbf{Definition 11: (Asymmetric class)} Every asymmetric variable in a group is an asymmetric class.

 \textbf{Definition 12: (Symmetric class)} A variable set $C_i=\{x_{i1},\cdots,x_{ik}\}$ and $k\in \{2,\cdots,n\}$ is a symmetric class if any arbitrary two variables in $C_i$ are symmetric.

  \textbf{Definition 13: (Independent class)} A variable set $C_i=\{x_{i1},\cdots,x_{ik}\}$ and $k\in \{1,\cdots,n-1\}$ is an independent class if every variable in $C_i$ is an independent variable.

\subsection{Compute Canonical Form}

The maximal canonical transformation $C_f$ is found after comparing all the candidate canonical transformations. We first try to use the $1^{st}$ DC signature value to group the variables. When the $1^{st}$ DC signature values can not resolve all the groups, we use the 2nd-order DC signature values and higher DC signature values until all the groups are resolved. Then, our algorithm generates a candidate canonical transformation.

Before search candidate canonical transformation, our algorithm need to do some  preparatory works including phase assignment, symmetry checking and independent variable checking. The method of determining the phase of Boolean function and variable is similar to \cite{Adbollahi2008}.

The first comparison values are $|f|$ and $|\overline f|$. If $|f|>|{\overline{f}}|$, there is no negation to output; otherwise, there is a negation to the output. When $|f|=|{\overline{f}}|$, our algorithm need to test $f$ and $\overline{f}$.

The phase of a variable is obtained by comparing the size of its positive and negative cofactors. The phase of $x_i$ is positive when $|f_{x_i}|>|f_{{\overline{x}}_i}|$, and negative when $|f_{x_i}|<|f_{{\overline{x}}_i}|$. When $|f_{x_i}|=|f_{{\overline{x}}_i}|$, we need to use a higher signature to determine its phase. If the higher signature cannot determine its phase, we must try both positive and negative.

If variable $x_i$ of a Boolean function $f$ is an independent variable, the phase of $x_i$ is assigned as positive when $|f_{x_i}|=|f_{{\overline{x}}_i}|$ because $x_i$ is independent to the Boolean function. If $f$ has $k$ independent variables and their phases cannot be determined by comparing the positive and negative cofactors, we may try two phases; which doubles the search space. Therefore, checking independent variables is a necessary operation.

\floatname{algorithm}{Procedure}
\renewcommand{\algorithmicrequire}{\textbf{Input:}}
\renewcommand{\algorithmicensure}{\textbf{Output:}}
    \begin{algorithm}
    \small
        \caption{Compute Canonical Form}
        \begin{algorithmic} 
            \Require $f$
            \Ensure  $\mathbb F$
            \Function {Canonical}{$f$}
            \State CT\_List=NULL
            \State Create Boolean function $f$ and compute $|f|$
            \If {$|f|<|\overline{f}|$}
                \State $f=\overline{f}$
            \EndIf
                \State Compute the $1^{st}$ DC signature of $f$
                \State Determine the phase of $x_1,\cdots,x_n$
                \State Check the independent variables
                \State Check the symmetric variable
                \State m=group($f$)
                \State SEARCH($f, CT\_list, group,m$)
           \If {$|f|=2^{n-1}$}
                \State $f=\overline{f}$
                \State Compute the $1^{st}$ difference signature of $f$
                \State Determine the phases of $x_1,\cdots,x_n$
                 \State Check the independent variables
                \State Check the symmetry variable
                \State m=group($f$)
                \State SEARCH($f, CT\_list, group,m$)
            \EndIf
             \State ${\cal T}={C_f}^{-1}$
             \State Return $f({\cal T}X)$
            \EndFunction
        \end{algorithmic}
    \end{algorithm}

The purpose of Procedure 1 is to compute the canonical form of a Boolean function $f$. The individual tasks in Procedure 1 are as follows.

1) Compute $|f|$ and determine the phase of $f$.

2) Compute the $1^{st}$ DC signature values of all variables of $f$.

3) The $1^{st}$ DC signature value determines the phases of all variables. Variables whose phases are not determined are handled in Procedure 2. The method used is the same as that described in \cite{Adbollahi2008}.

4) Check the symmetry of all variables.

5) Check the independent variables. If the phases of independent variables are not determined, assign the positive phase to them.

6) Group all variables by comparing their $1^{st}$ DC signature values.

7) Call Procedure 2 to get the maximal canonical transformation $C_f$.

8) Compute canonical form by the maximal canonical transformation $C_f$.

\textbf{Example 1:} Consider a 7-input Boolean function $f(X)=x_3{\overline{x}}_5x_7+{\overline{x}}_1{\overline{x}}_2x_4+{\overline{x}}_1{\overline{x}}_2{\overline{x}}_4x_5{\overline{x}}_6+
x_3{\overline{x}}_4x_5{\overline{x}}_6+x_2{\overline{x}}_3{\overline{x}}_4x_5{\overline{x}}_6+{\overline{x}}_2{\overline{x}}_3{\overline{x}}_4x_5{\overline{x}}_6  $, whose phase assignment and group variables are as follows.

Procedure 1 computes $|f|=46$ and assigns a positive phase to $f$. Then, it computes the $1^{st}$ DC signature value of all the variables of $f$, and the results are \{(16,28), (16,28), (30,28), (22,44), (24,44), (15,32), (30,28)\}. The variables $x_3$, $x_5$ and $x_7$ are positive, and the others are negative.

Check the symmetry of every variable. There are two symmetric class $\{x_1,x_2\}$ and $\{x_3,x_7\}$. Group the variables according to the $1^{st}$ DC signature value. The grouping result is $G=\{G_1,G_2,G_3\}$, and $G_1=\{C_{11}\}=\{{\overline{x}}_6\}$, $G_2=\{C_{21},C_{22}\}=\{\{{\overline{x}}_1,{\overline{x}}_2\},\{x_3,x_7\}\}$ and $G_3=\{C_{31},C_{32}\}=\{\{{\overline{x}}_4\},\{x_5\}\}$.

\subsection{Searching the Canonical Transformations}

After obtaining the initial group results, Procedure 2 begins to search candidate canonical transformations. Procedure 2 first addresses the first group $G_1$, then group $G_2$, and so on until all groups have been resolved. A group $G_i$ is resolved when $G_i$ has only one class. There are three possible cases when group $G_i$ is resolved.

1) Group $G_i$ has only one asymmetric variable and the phase of this variable is determined.

2) Group $G_i$ has multiple variables, an arbitrary two variables of $G_i$ are symmetric, and the phases of all variables of $G_i$ are determined.

3) Group $G_i$ has one or more variables, all of which are independent variables.

Because the Boolean difference signature of an independent variable is $0$ and the Boolean difference signature of dependent variable must not be $0$, a group contains only independent variables when any independent variable occurs.

   When searching candidate canonical formations, all the candidate canonical transformations are stored in a tree and Procedure 2 uses a depth-first search to find candidate canonical transformations. Each branch of the candidate canonical transformation tree is a candidate canonical transformation. In Procedure 2, CT\_list is the candidate canonical transformation tree, which is initialized to NULL. $T$ is the candidate canonical transformation found in the search process, and $C_f$ is the maximal canonical transformation, which is initialized empty.

\floatname{algorithm}{Procedure}
\renewcommand{\algorithmicrequire}{\textbf{Input:}}
\renewcommand{\algorithmicensure}{\textbf{Output:}}
    \begin{algorithm}
    \small
        \caption{Canonical transformation search process}
        \begin{algorithmic} 
            \Require $f,map\_list, group,m$
            \Ensure  $C_f$
            \Function {Search}{$f,CT\_list,group,m$}
                 \If {$D_1$}   
                     \If {$Empty(C_f)$}
                        \State $C_f=T$
                     \ElsIf{$D^{C_f} \prec D^T $}
                         \State $C_f=T$
                      \EndIf
                 \Else 
                     \ForAll {$G_i,i \in (1,\cdots,m)$}
                        \If {$D_2$} 
                         \State  break
                        \EndIf
                     \EndFor

                     \If {$D_3$}  
                         \State Add the variables in $G_i$ to CT\_list
                         \State SEARCH($f, CT\_list, group,m$)
                     \Else
                       \ForAll{$C_{ij},C_{ij}\in G_i$ } 
                            \State Split $G_i$
                            \State create node\_list
                       \EndFor
                       \ForAll{$node\_j$ in node\_list}
                         \State UPDATE\_SEQUENCE($node\_j,group$)
                         \State m=m+1
                         \If {$D_1$}   
                             \If {$Empty(C_f)$}
                                 \State $C_f=T$
                              \ElsIf{$D^{C_f} \prec D^T $}
                                       \State $C_f=T$
                               \EndIf
                         \Else
                           \State Add the variables in $G_i$ to CT\_list
                           \State Update\_signature($f$)
                           \State m=group($f$)
                           \State SEARCH($f, CT\_list, group,m$)
                         \EndIf
                      \EndFor
                \EndIf
                \EndIf
            \EndFunction
        \end{algorithmic}
    \end{algorithm}

Let $G=\{G_1,G_2,\cdots,G_m\}$ be the grouping results that Procedure 1 obtained using the $1^{st}$ DC signature values. The first step in Procedure 2 is to check whether all groups are resolved. Here we use the condition $D_1$ to denote whether all groups are resolved.

When $D_1$ is true, Procedure 2 finds a candidate canonical transformation $T$. When this occurs, Procedure 2 assigns $T$ to $C_f$ if $C_f$ is empty or $D^{C_f} \prec D^T $. Then, Procedure 2 continues to search for other candidate canonical transformations until all the candidate canonical transformations have been searched.

When $D_1$ is false, some groups could not be resolved. At this point, Procedure 2 searches the group $G_i$ which is an unresolved group that has the minimal sequence number. Condition $D_2$ becomes true when an unresolved group is found.

When group $G_i$ (the unresolved group having the minimal sequence number) is found, it is handled in two ways.

1) Group $G_i$ can be resolved.

The condition $D_3$ denotes whether group $G_i$ can be resolved. When $D_3$ is true, Procedure 2 adds all the variables in $G_i$ to the candidate canonical transformation tree and calls itself recursively.

In Example 1, the condition $D_1$ is false. Procedure 2 finds that group $G_1$ is the first unresolved group and that it can be resolved in this call. Procedure 2 adds variable ${\overline{x}}_6$ to the candidate canonical transformation tree, giving a new layer with a node. Then, Procedure 2 calls itself recursively.

2) Group $G_i$ cannot be resolved.

When condition $D_3$ is false, Procedure 2 splits group $G_i$. Group $G_i$ has $k$ classes; thus, there are $k$ approaches for splitting it. Let group $G_i=\{C_{i1},\cdots,C_{ik}\}$, select one class as group $G_i$; the other classes are in group $G_{i+1}$. The first splitting approach is $P_1=\{G_i,G_{i+1}\}=\{\{C_{i1}\},\{C_{i2},\cdots,C_{ik}\}\}$, the second splitting approach is $P_2=\{G_i,G_{i+1}\}=\{\{C_{i2}\},\{C_{i1},C_{i3},\cdots,C_{ik}\}\}$, and the \emph{k}th splitting approach is $P_k=\{G_i,G_{i+1}\}=\{\{C_{ik}\},\{C_{i1},\cdots,C_{i(k-1)}\}\}$. Note when the phases of the variables of group $G_i$ are not determined, there are $2k$ splitting approaches, because we need to try the positive and the negative.

Suppose there are $k$ splitting approaches. These $k$ splitting approaches are stored in the variable node\_list. Then, Procedure 2 selects one splitting approach from the variable node\_list in order. Let Procedure 2 selects the \emph{q}th splitting approach $P_q=\{G_i,G_{i+1}\}=\{\{C_{iq}\},\{C_{i1},\cdots,C_{i(q-1)},C_{i(q+1)},\cdots,C_{ik}\}\}$, Procedure 2 calls UPDATE\_SEQUENCE() to update the sequence number of the groups from $G_i$ to $G_m$. the group sequence number of the variables of $C_{iq}$ remains unchanged and the group sequence number of the variables of $\{C_{i1},\cdots,C_{i(q-1)},C_{i(q+1)},\cdots,C_{ik}\}$ is incremented by one. Finally, the group sequence number of the variables of original group $G_{i+1},\cdots, G_m$ is incremented by one, and the value of $m$ is incremented by one.

Then, Procedure 2 checks the condition $D_1$. When condition $D_1$ is true, Procedure 2 creates a candidate canonical transformation $T$ and compares the DC signature vector of $T$ and $C_f$. When condition $D_1$ is false, the \emph{q}th branch is added to the $|C_{iq}|$ layers as one layer with one node. In other words, $|C_{iq}|$ expresses the number of variables in $C_{iq}$. Procedure 2 calls Update\_signature() to update the DC signature value of $f$ and regroups the variables of $G_{i+1},\cdots,G_m$, and calls itself recursively.

Procedure 2 starts with the first splitting approach and then processes the next splitting approach until all splitting approaches have been processed.

 In Example 1, after handling group $G_1$, group $G_2$ is the next unresolved group with the minimal sequence number. Group $G_2$ has two symmetric classes; therefore, it can not be resolved. There are two ways to split group $G_2$: $P_1=\{G_2,G_3\}=\{\{{\overline{x}}_1,{\overline{x}}_2\},\{x_3,x_7\}\}$ and $P_2=\{G_2,G_3\}=\{\{x_3,x_7\},\{{\overline{x}}_1,{\overline{x}}_2\}\}$. The first branch added to the candidate canonical transformation tree has two layers with the nodes ${\overline{x}}_1$ and ${\overline{x}}_2$. The second branch added to the candidate canonical transformation tree has two layers with the nodes $x_3$ and $x_7$.

  Procedure 2 selects one splitting approach in sequence and updates the DC signature values. The method for updating the DC signature is similar to that used in \cite{Adbollahi2008} to update signatures. In Example 1, Procedure 2 uses variable ${\overline{x}}_6$ to compute the 2nd-order DC signature value for the other variables. Thus, there are new DC signature values, and Procedure 2 regroups all the unresolved groups and calls itself recursively.

Our algorithm uses three strategies to reduce the search space.

1) A symmetric class has multiple variables. When we group and sort variables, all the variables of a symmetric class are regarded as a single operational object.

2) Take advantage of the independence of independent variable. When the phase of independent variables cannot be determined, we assign positive phase to them. If we do not make use of independent variable, a independent class $C_{il}$ is a symmetric class. When the phases of the variables in $C_{il}$ are not determined, the number of candidate canonical transformations is doubled. When there are other symmetric classes having the same signatures to $C_{il}$, the number of candidate canonical transformations is grow even more.

3) The combination of the cofactor signature and Boolean difference signature can better distinguish variables.

In Example 1, after updating the DC signature using variable ${\overline{x}}_1$, variables ${\overline{x}}_4$ and $x_5$ are also resolved. There are two candidate canonical transformations $T1=\{{\overline{x}}_6,{\overline{x}}_1,{\overline{x}}_2,x_3,x_7,{\overline{x}}_4,x_5\}$ and $T2=\{{\overline{x}}_6,x_3,x_7,{\overline{x}}_1,{\overline{x}}_2,{\overline{x}}_4,x_5\}$. The candidate canonical transformation tree of Example 1 is shown in Fig. 1.

\begin{figure}[!h]
\centering
\includegraphics[width=0.25\textwidth,height=5cm]{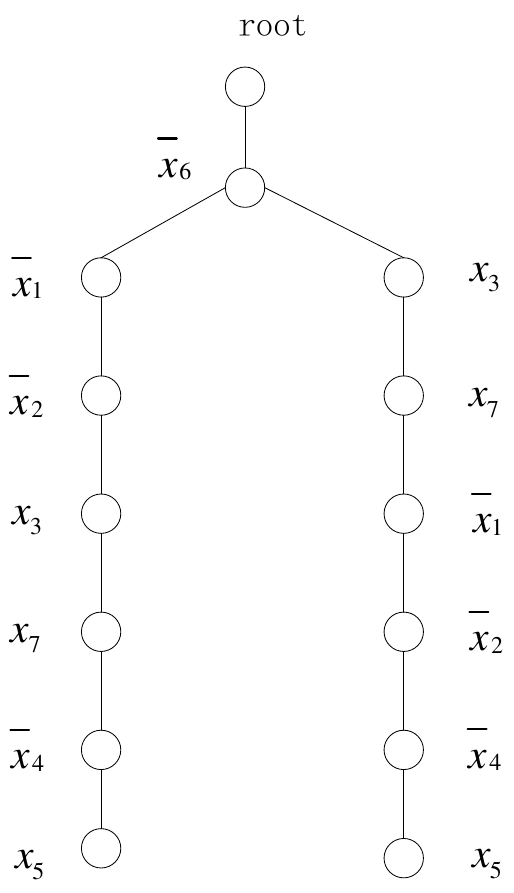}
\caption{The candidate canonical transformation tree of Example 1}
\label{fig.1}
\end{figure}

For a 7-input Boolean function, there are $7!2^7$ NP transformations, while only two candidate canonical transformations exist after using our algorithm in Example 1. Through the DC signature vector comparison, the maximal canonical transformation $C_f$ is $\{{\overline{x}}_6,{\overline{x}}_1,{\overline{x}}_2,x_3,x_7,{\overline{x}}_4,x_5\}$.

To demonstrate that our algorithm is superior to the algorithm proposed in \cite{Adbollahi2008}, we use both algorithms to search the maximal candidate canonical transformations in Example 2.

\textbf{Example 2:} Consider a 6-input Boolean function $f(X)={\overline{x}}_1x_2x_3{\overline{x}}_4+x_1x_2{\overline{x}}_3{\overline{x}}_4+{\overline{x}}_1{\overline{x}}_2{\overline{x}}_3x_4+{\overline{x}}_1x_2x_3x_6+
{\overline{x}}_1x_2x_4x_5{\overline{x}}_6+x_1{\overline{x}}_2x_3{\overline{x}}_6+x_1x_2{\overline{x}}_3x_6+{\overline{x}}_1{\overline{x}}_2{\overline{x}}_3{\overline{x}}_4{\overline{x}}_5+
{\overline{x}}_1{\overline{x}}_2x_3{\overline{x}}_4x_6+{\overline{x}}_1{\overline{x}}_2x_3x_5x_6+{\overline{x}}_1{\overline{x}}_2{\overline{x}}_3x_5{\overline{x}}_6+
{\overline{x}}_1x_2{\overline{x}}_3x_4{\overline{x}}_5{\overline{x}}_6+x_1x_2x_3x_4{\overline{x}}_5{\overline{x}}_6+x_1{\overline{x}}_2{\overline{x}}_3{\overline{x}}_4x_5x_6     $. We use both the signature vector from \cite{Adbollahi2008} and the DC signature vector proposed in this paper to search the candidate canonical transformations.

1) Using the algorithm of \cite{Adbollahi2008}, the candidate canonical transformation search process is as follows.

The algorithm from \cite{Adbollahi2008} computes the $1^{st}$ signature, checks the symmetry and groups variables. The Boolean function $f$ has 32 minterms, i.e., $|f|=32$. The results of the $1^{st}$ signature are $|f_{x_1}|=13$ and $|f_{x_2}|=|f_{x_3}|=|f_{x_4}|=|f_{x_5}|=|f_{x_6}|=16$. There are no symmetric classes. The phase of variable $x_1$ is negative and the phases of the other variables are not determined. According to the $1^{st}$ signatures, the grouping result is $G=\{G_1,G_2\}=\{\{x_1\},\{x_2,x_3,x_4,x_5,x_6\}\}$.

 The first group $G_1$ can be resolved, and the variable ${\overline{x}}_1$ is used to compute the 2nd-order signature of the variables in $G_2$. The results are $|f_{{\overline{x}}_1x_2}|=9$ and $|f_{{\overline{x}}_1x_3}|=|f_{{\overline{x}}_1x_4}||f_{{\overline{x}}_1x_5}||f_{{\overline{x}}_1x_6}|=10$. We know that $|f_{{\overline{x}}_1}|=|f|-|f_{x_1}|=19$; therefore, we can determine the phases of these 5 variables. Variable $x_2$ is negative and the others are positive. But $G_2$ cannot be resolved: these 5 variables have the same 2nd-order signature. Group $G_2$ has \emph{5} asymmetric classes, the algorithm of \cite{Adbollahi2008} splits $G_2$ and there are five ways to split it.

\begin{itemize}
  \item $P_1=\{G_2,G_3\}=\{\{x_2\},\{x_3,x_4,x_5,x_6\}\}$
  \item $P_2=\{G_2,G_3\}=\{\{x_3\},\{x_2,x_4,x_5,x_6\}\}$
  \item $P_3=\{G_2,G_3\}=\{\{x_4\},\{x_2,x_3,x_5,x_6\}\}$
  \item $P_4=\{G_2,G_3\}=\{\{x_5\},\{x_2,x_3,x_4,x_6\}\}$
  \item $P_5=\{G_2,G_3\}=\{\{x_6\},\{x_2,x_3,x_4,x_5\}\}$
\end{itemize}

The algorithm of \cite{Adbollahi2008} traverses each possible split shown above. We select the first split approach as an example. The updated groups are $G_1=\{x_1\}$, $G_2=\{x_2\}$ and $G_3=\{x_3,x_4,x_5,x_6\}$.

 Group $G_2$ can be resolved; therefore, we use variable ${\overline{x}}_2$ to compute 2nd-order signature and group variables. The 2nd-order signatures of the unresolved variables are $|f_{{\overline{x}}_2x_3}|=|f_{{\overline{x}}_2x_4}|=|f_{{\overline{x}}_2x_5}|=|f_{{\overline{x}}_2x_6}|=8$. From these results, we also know that group $G_3$ also cannot be resolved. There are four ways to split it as follows.

\begin{itemize}
  \item $P_1=\{G_3,G_4\}=\{\{x_3\},\{x_4,x_5,x_6\}\}$
  \item $P_2=\{G_3,G_4\}=\{\{x_4\},\{x_3,x_5,x_6\}\}$
  \item $P_3=\{G_3,G_4\}=\{\{x_5\},\{x_3,x_4,x_6\}\}$
  \item $P_4=\{G_3,G_4\}=\{\{x_6\},\{x_3,x_4,x_5\}\}$
\end{itemize}

Continuing, the algorithm of \cite{Adbollahi2008} computes its signature and splits the group. Because variables $x_2,x_3,x_4,x_5$ and $x_6$ always have the same 2nd-order signature, the following computations will generate a total of $5\times 4 \times 3 \times 2$ candidate canonical transformations.

2) Using the algorithm proposed by this paper, the candidate canonical transformation search process is as follows.

 Our algorithm computes the $1^{st}$ DC signature value of each variable, checks the symmetry, and searches the independent variables and groups variables by the $1^{st}$ DC signature value.

    Boolean function $f$ has no symmetry classes or independent classes. The $1^{st}$ DC signature values are $(|f_{x_1}|,|f_{x_1}^{'}|)=(13,64)$, $(|f_{x_2}|,|f_{x_2}^{'}|)=(16,36)$, $(|f_{x_3}|,|f_{x_3}^{'}|)=(16,52)$, $(|f_{x_4}|,|f_{x_4}^{'}|)=(16,20)$, $(|f_{x_5}|,|f_{x_5}^{'}|)=(16,12)$ and $(|f_{x_6}|,|f_{x_6}^{'}|)=(16,28)$.

    Because $|f|=32$, we can determine that the phase of variable $x_1$ is negative and the phases of others are not determined. Our algorithm groups the variables and generates 6 groups: $G_1=\{x_1\}$, $G_2=\{x_3\}$, $G_3=\{x_2\}$,$G_4=\{x_6\}$, $G_5=\{x_4\}$ and $G_6=\{x_5\}$.

Group $G_1$ can be resolved. The other groups are not resolved because the phases of the variables in these groups have not been determined. Our algorithm uses the variable ${\overline{x}}_1$ to compute the 2nd-order DC signature values, which are: $(|f_{{\overline{x}}_1x_2}|,|f_{{\overline{x}}_1x_2}|^{'})=(9,18)$, $(|f_{{\overline{x}}_1x_3}|,|f_{{\overline{x}}_1x_3}|^{'})=(10,26)$, $(|f_{{\overline{x}}_1x_4}|,|f_{{\overline{x}}_1x_4}|^{'})=(10,10)$, $(|f_{{\overline{x}}_1x_5}|,|f_{{\overline{x}}_1x_5}|^{'})=(10,6)$, and $(|f_{{\overline{x}}_1x_6}|,|f_{{\overline{x}}_1x_6}|^{'})=(10,14)$.

From the above 2nd-order DC signature values, the phases of all variables are determined. Therefore, groups $G_2$, $G_3$, $G_4$, $G_5$ and $G_6$ are all resolved. One candidate canonical transformation is generated: $\{{\overline{x}}_1,x_3,{\overline{x}}_2,x_6,x_4,x_5\}$.

Because $|f|=|{\overline{f}}|$, the algorithm of \cite{Adbollahi2008} and ours must both perform the same search for $\overline{f}$. However, the Boolean function $f$ results in 240 candidate canonical transformations when computed by the algorithm of \cite{Adbollahi2008} but only 2 when computed by the algorithm proposed in this paper.

Using our algorithm, the NPN Boolean matching algorithm is denoted as follows: Given two Boolean function $f(X)$ and g(X), Boolean function $f$ is NPN-equivalent to $g$ if the canonical form $\mathbb F$ of $f$ is equal to the canonical form $\mathbb G$ of $g$.

\section{Experimental Results}

We reimplemented the algorithm of \cite{Adbollahi2008} to compare it with the algorithm of this paper and obtain supporting experimental evidence. The two algorithms were tested on a set of randomly generated circuits and a set of MCNC benchmark circuits. The two circuit sets include a number of NPN-equivalent circuits. We tested the NPN matching runtime and the number of candidate canonical transformations generated when searching for the canonical transformation of a Boolean function. The runtime is measured in seconds. The following experimental results were obtained on a computer equipped with 3.3 GHz CPU and 4GB RAM.

 Tables \uppercase\expandafter{\romannumeral1} and \uppercase\expandafter{\romannumeral2} show the experimental results on the random circuit and MCNC benchmark circuit sets, respectively. In these two tables, the first column shows the number of inputs (\#I), the second and third columns show the average matching runtime (\#A.T) and the average number of candidate canonical transformations (\#A.C.T) of the algorithm in \cite{Adbollahi2008}, respectively, and the forth and fifth columns show the average matching runtime and the average number of candidate canonical transformations of our algorithm, respectively.

\begin{table}[htp]
\label{Tab1}
\caption{Boolean matching results on the random circuits}
\centering
\begin{tabular}{|p{0.3cm}|p{1cm}| p{1cm}|p{1cm}|p{1cm}|}
\hline
\#I &\#A.T of \cite{Adbollahi2008} &\# A.C.N of \cite{Adbollahi2008} &\# A.T of ours  & \# A.C.N of ours\\
\hline
7&0.0007 	&13.2	&0.0005 	&7.3 	  \\
\hline
8&0.0009	&22.1 	&0.0004 	&5.3  \\
\hline
9 &0.0010 	&16.8 	&0.0007 	&6.1 \\
\hline
10 &0.0013	&19.6 	&0.0004 	&3.6  \\
\hline
11&0.0032	&19.1	&0.0020	&4.3 \\
\hline
12& 0.0068 	&12.5 	&0.0048 	&4.5  \\
\hline
13 & 0.0099	&17.2 	&0.0043 	&3.7   \\
 \hline
14 & 0.0097 	&14.8 	&0.0062 	&4.2  \\
\hline
15 & 0.0078 	&16.5 	&0.0044	&4.2   \\
\hline
16& 0.0249 	&13.9	&0.0135 	&5.0   \\
\hline
17 & 0.0326 	&23.4	&0.0208 &5.9  \\
\hline
18 & 0.0423 	&13.3 	&0.0261 	&5.0  \\
\hline
19 & 0.1565 	&15.8	&0.1070	&5.8  \\
\hline
20 & 0.1195	&12.5 	&0.0974 	&5.6  \\
\hline
21 & 0.0859 	&13.3 	&0.0708 	&4.9   \\
\hline
22& 0.1349 	&18.3	&0.1186	&6.8  \\
\hline
\end{tabular}
\end{table}

The experimental results listed in Table \uppercase\expandafter{\romannumeral1} show that our algorithm improves the speed of the Boolean matching process and reduces the search space effectively. In the best case, our algorithm's runtime is 70\% faster and its search space is 82\% smaller than that of the algorithm from \cite{Adbollahi2008}. On average, our algorithm improves the runtime by 37\% and reduces the search space by 67\%. Fig. 2 shows the results on the equivalent random circuits.

\begin{figure}[!h]
\centering
\includegraphics[width=0.5\textwidth,height=9cm]{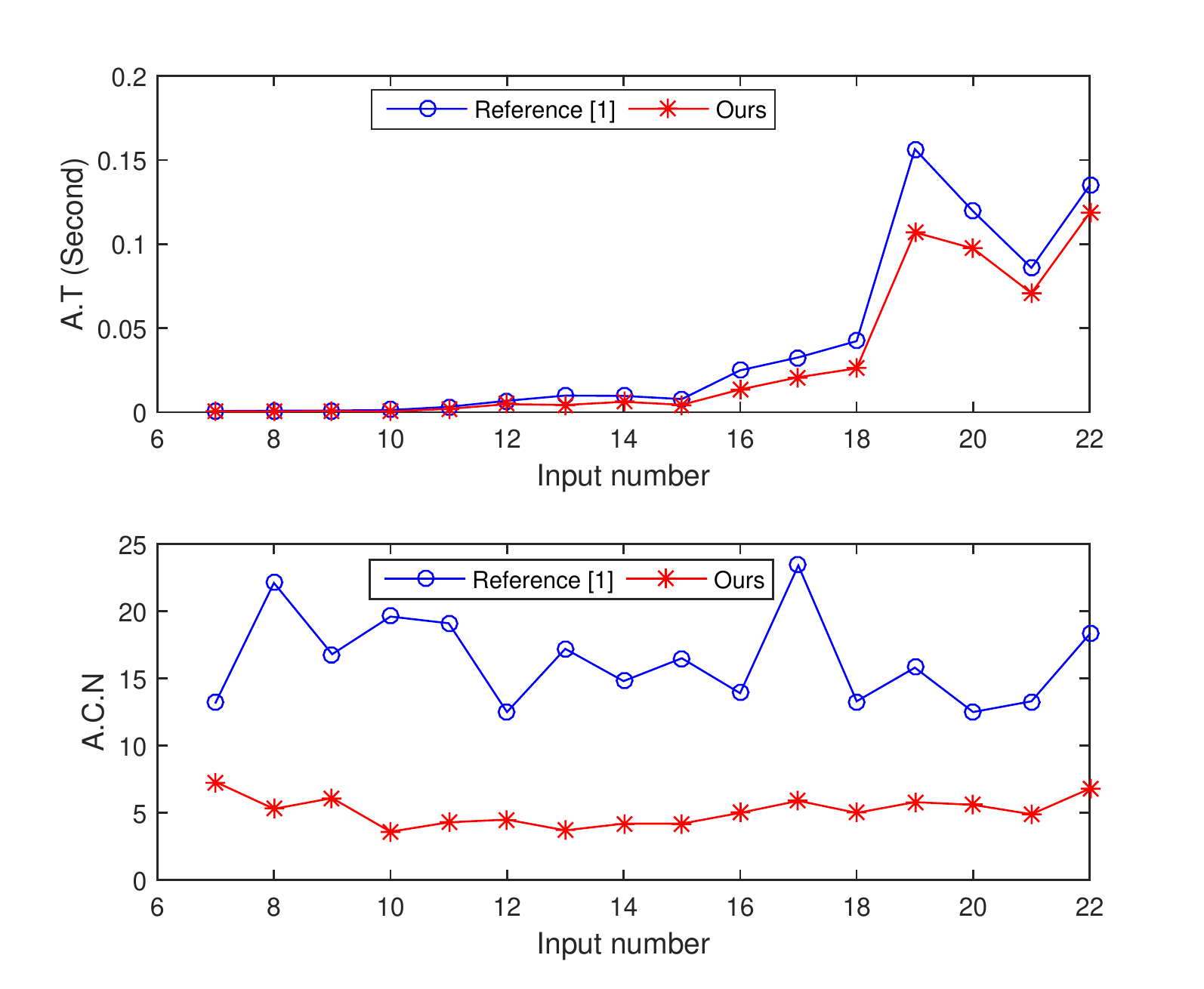}
\caption{The comparison results on the equivalent random circuits}
\label{fig.1}
\end{figure}
\footnotetext[1]{A.T: Average runtime, A.C.N: Average number of candidate canonical transformations}

\begin{table}[htp]
\label{Tab2}
\caption{Boolean matching results on the MCNC benchmark circuits}
\centering
\begin{tabular}{|p{0.3cm}|p{1cm}| p{1cm}|p{1cm}|p{1cm}|}
\hline
\#I &\#A.T of \cite{Adbollahi2008} &\# A.C.N of \cite{Adbollahi2008} &\# A.T of ours  & \# A.C.N of ours\\
\hline
7&0.00215 	&66.7 	&0.00037 	&2.4 \\
\hline
8&0.00989 	&27.1 	&0.00939 	&26.3 \\
\hline
9 &0.00163 	&3.0 	&0.00169 	&2.8\\
\hline
10 &0.00098 	&2.1 	&0.00103 	&2.1 \\
\hline
11&0.00283 	&3.2 	&0.00211	&2.3\\
\hline
12& 0.00177 	&2.7 	&0.00173 	&2.2 \\
\hline
13 & 0.04242 	&7.8 	&0.02971 	&6.5  \\
\hline
14 & 0.00713 	&2.1 	&0.00784 	&2.0 \\
\hline
15 & 0.02730 	&2.4 	&0.02818	&2.4  \\
\hline
16& 0.00238 	&2.9	&0.00246 	&2.4  \\
\hline
17 & 0.18872 	&2.8	&0.19304 	&2.5 \\
\hline
18 & 0.21707 	&4.4 	&0.21725 	&2.4 \\
\hline
19 & 1.13960 	&3.4	&1.04902	&2.1 \\
\hline
20 & 1.90159 	&3.8 	&2.07058 	&3.7 \\
\hline
21 & 7.1248 	&2.0 	&6.6848 	&1.8  \\
\hline
22& 13.9547 	&4.5 	&14.3204 	&4.0 \\
\hline
\end{tabular}
\end{table}

Form Table \uppercase\expandafter{\romannumeral2}, we can see that the average runtime improves by 83\% and the search space is reduced by 96\% compared with \cite{Adbollahi2008} when the number of input variables is 7. However, neither the runtime nor the search space are drastically improved in the other input cases. In majority of the circuits of the MCNC benchmark, using only cofactor signature serves to quickly identify the variables. Therefore, the average runtime of the algorithm of \cite{Adbollahi2008} when tested on 22 inputs is slightly faster than ours. Despite this, our algorithm improves the average runtime by 6\% and reduces the average search space by 19\% compared with that of \cite{Adbollahi2008} when tested on the MCNC benchmark circuits.

\section{Conclusions}

In this paper, we propose a canonical-based Boolean matching algorithm. The proposed DC signature vector is more effective in computing canonical form than the algorithm of \cite{Adbollahi2008}. We test the runtime and the canonical transformation search space of Boolean matching algorithms with 7-22 inputs. The experimental results show that the average runtime of our algorithm is 37\% faster and its search space is 67\% smaller than those of the algorithm proposed in \cite{Adbollahi2008}. The algorithm proposed in this paper is highly effective at reducing the search space and enhancing the Boolean matching speed. In future work, we plan to extend this Boolean matching approach to Boolean matching with don't care set and multiple-output Boolean matching.

\ifCLASSOPTIONcaptionsoff
  \newpage
\fi



%

\section*{Acknowledgment}

We would like to thank the National Natural Science Foundation of China (Grant No. 61572109, No. 11371003) and the Special Fund for Bagui Scholars of Guangxi (Grant No. 113000200230010) for their support for technology.

\bibliographystyle{./IEEEtran}

\bibliography{./IEEEabrv,./paper}

\begin{thebibliography}{10}
\providecommand{\url}[1]{#1}
\csname url@samestyle\endcsname
\providecommand{\newblock}{\relax}
\providecommand{\bibinfo}[2]{#2}
\providecommand{\BIBentrySTDinterwordspacing}{\spaceskip=0pt\relax}
\providecommand{\BIBentryALTinterwordstretchfactor}{4}
\providecommand{\BIBentryALTinterwordspacing}{\spaceskip=\fontdimen2\font plus
\BIBentryALTinterwordstretchfactor\fontdimen3\font minus
  \fontdimen4\font\relax}
\providecommand{\BIBforeignlanguage}[2]{{%
\expandafter\ifx\csname l@#1\endcsname\relax
\typeout{** WARNING: IEEEtran.bst: No hyphenation pattern has been}%
\typeout{** loaded for the language `#1'. Using the pattern for}%
\typeout{** the default language instead.}%
\else
\language=\csname l@#1\endcsname
\fi
#2}}
\providecommand{\BIBdecl}{\relax}
\BIBdecl

\bibitem{Adbollahi2008}
A.~Adbollahi and M.~Pedram, ``Symmetry detection and boolean matching utilizing
  a signature-based canonical form of boolean functions,'' \emph{IEEE
  Transactions on Computer-Aided Design of Integrated Circuits and Systems},
  vol.~27, no.~6, pp. 1128--1137, 2008.

\bibitem{YanZhang2016}
Y.~Zhang, G.~Yang, W.~N. Hung, and J.~Zhang, ``Computing affine equivalence
  classes of boolean functions by group isomorphism,'' \emph{IEEE Transactions
  on Computers}, vol.~65, no.~12, pp. 3606--3616, 2016.

\bibitem{B.Kapoor11995}
B.~Kapoor, ``Improved technology mapping using a new approach to boolean
  matching,'' in \emph{European Design and Test Conference,Proceedings}, Paris,
  France, Mar. 1995, pp. 86--90.

\bibitem{Mailhot1990}
F.~Mailhot and G.~De~Micheli, ``Technology mapping using boolean matching and
  don't care sets,'' in \emph{Design Automation Conference, 1990., EDAC.
  Proceedings of the European}, Glasgow, UK, 1990, pp. 212--216.

\bibitem{M.Damiani2003}
M.~Damiani and A.~Y. Selchenko, ``Boolean technology mapping based on logic
  decomposition,'' in \emph{Integrated Circuits and Systems Design, 2003. SBCCI
  2003. Proceedings. 16th Symposium on}, Sao Paulo, Brazil, Sep. 2003, pp.
  35--40.

\bibitem{Debnath2004}
D.~Debnath and T.~Sasao, ``Efficient computation of canonical form for boolean
  matching in large libraries,'' in \emph{Design Automation Conference, 2004.
  Proceedings of the ASP-DAC 2004. Asia and South Pacific}, Yohohama, Japan,
  Jan. 2004, pp. 591--596.

\bibitem{Agosta2009}
G.~Agosta, F.~Bruschi, G.~Pelosi, and D.~Sciuto, ``A transform-parametric
  approach to boolean matching,'' \emph{IEEE Transactions on Computer-Aided
  Design of Integrated Circuits and Systems}, vol.~28, no.~6, pp. 805--817,
  2009.

\bibitem{Chai2006}
D.~Chai and A.~Kuehlmann, ``Building a better boolean matcher and symmetry
  detector,'' in \emph{Design, Automation and Test in Europe, 2006. DATE '06.
  Proceedings}, Munich, Germany, 2006, pp. 1--6.

\bibitem{Ciric2003}
J.~Ciric and C.~Sechen, ``Efficient canonical form for boolean matching of
  complex functions in large libraries,'' \emph{IEEE Transactions on
  Computer-Aided Design of Integrated Circuits and Systems}, vol.~22, no.~5,
  pp. 535--544, 2003.

\bibitem{Agosta2007}
G.~Agosta, F.~Bruschi, G.~Pelosi, and D.~Sciuto, ``A unified approach to
  canonical form-based boolean matching,'' in \emph{Design Automation
  Conference, 2007. DAC '07. 44th ACM/IEEE}, San Diego, CA, 2007, pp. 841--846.

\bibitem{Huang2013}
Z.~Huang, L.~Wang, and Y.~Nasikovskiy, ``Fast boolean matching based on npn
  classification,'' in \emph{Field-Programmable Technology (FPT), 2013
  International Conference on}, Kyoto, Japan, Dec. 2013, pp. 310--313.

\bibitem{Petkovska2016}
A.~Petkovska, M.~Soeken, G.~D. Micheli, P.~Ienne, and A.~Mishchenko, ``Fast
  hierarchical npn classification,'' in \emph{International Conference on Field
  Programmable Logic and Applications}, Lausanne, Switzerland, Dec. 2016, pp.
  1--4.

\bibitem{K.C.Chen1993}
K.~C. Chen and J.~C.~Y. Yang, ``Boolean matching algorithms,'' in
  \emph{International Symposium on VLSI Technology, Systems, and Applications,
  Proceedings}, Taipei, Taiwan, May 1993, pp. 44--48.

\bibitem{Abdollahi2008}
A.~Abdollahi, ``Signature based boolean matching in the presence of don't
  cares,'' in \emph{Design Automation Conference, 2008. DAC 2008. 45th
  ACM/IEEE}, Anaheim, CA, 2008, pp. 642--647.

\bibitem{Y.T.Lai1992}
Y.~T. Lai, S.~Sastry, and M.~Pedram, ``Boolean matching using binary decision
  diagrams with applications to logic synthesis and verification,'' in
  \emph{Computer Design: VLSI in Computers and Processors, 1992. ICCD'92.
  Proceedings, IEEE 1992 International Conference on}, Cambridge,MA, Oct. 1992,
  pp. 452--458.

\bibitem{Juling}
J.~Zhang, G.~Yang, W.~N. Hung, and Y.~Zhang, ``An efficient npn boolean
  matching algorithm based on structural signature and shannon expansion.''

\bibitem{WangKuoHua2009}
K.~H. Wang, C.~M. Chan, and J.~C. Liu, ``Simulation and sat-based boolean
  matching for large boolean networks,'' in \emph{Design Automation Conference,
  2009. DAC '09. ACM/IEEE}, San Francisco, CA, 2009, pp. 396--401.

\bibitem{Katebi2010}
Katebi, Hadi, and L.~Igor, ``Large-scale boolean matching,'' \emph{Advanced
  Techniques in Logic Synthesis Optimizations Applications}, pp. 771--776,
  2010.

\bibitem{Matsunaga2016}
Y.~Matsunaga, ``Accelerating sat-based boolean matching for heterogeneous fpgas
  using one-hot encoding and cegar technique,'' in \emph{Design Automation
  Conference}, Chiba, Japan, Jan. 2016, pp. 255--260.

\bibitem{WangXiuQin2013}
X.~Q. Wang and Y.~Yang, ``New approach of exploiting symmetry in sat-based
  boolean matching for fpga technology mapping,'' in \emph{IEEE International
  Conference on Vehicular Electronics and Safety}, Dongguan, China, 2013, pp.
  282--285.

\bibitem{Safarpour2006}
S.~Safarpour, A.~Veneris, G.~Baeckler, and R.~Yuan, ``Efficient sat-based
  boolean matching for fpga technology mapping,'' in \emph{Design Automation
  Conference, 2006 ACM/IEEE}, San Francisco, CA, 2006, pp. 466--471.

\bibitem{KuoHua2007}
K.~H. Wang and C.~M. Chan, ``Incremental learning approach and sat model for
  boolean matching with don¡¯t cares,'' in \emph{Ieee/acm International
  Conference on Computer-Aided Design}, San Jose, CA, Nov. 2007, pp. 234--239.

\bibitem{Moore2008}
J.~Moore, K.~Fazel, M.~A. Thornton, and D.~M. Miller, ``Boolean function
  matching using walsh spectral decision diagrams,'' in \emph{Design,
  Applications, Integration and Software, 2006 IEEE Dallas/CAS Workshop on},
  Richardson, TX, Oct. 2008, pp. 127--130.

\bibitem{Chih-Fan2010}
C.-F. Lai, J.-H.~R. Jiang, and K.-H. Wang, ``Boolean matching of function
  vectors with strengthened learning,'' in \emph{Computer-Aided Design (ICCAD),
  2010 IEEE/ACM International Conference on}, San Jose, CA, Nov. 2010, pp.
  596--601.

\bibitem{Hinsberger1998}
U.~Hinsberger and R.~Kolla, ``Boolean matching for large libraries,'' in
  \emph{Design Automation Conference}, San Francisco, CA, May 1998, pp.
  206--211.

\bibitem{Jin2006}
J.~S.Zhang, M.~Chrzanowska-Jeske, A.~Mishchenko, and J.~R. Burch, ``Linear
  cofactor relationships in boolean functions,'' \emph{IEEE Transactions on
  Computer-Aided Design of Integrated Circuits and Systems}, vol.~25, no.~6,
  pp. 1011--1023, 2006.

\end{thebibliography}

\end{document}